\documentclass[conference]{IEEEtran}

\IEEEoverridecommandlockouts
\usepackage{cite}
\usepackage{amsmath,amssymb,amsfonts}
\usepackage{algorithmic}
\usepackage{graphicx}
\usepackage{textcomp}
\usepackage{xcolor}
\usepackage{algorithmic}
\usepackage{algorithm}
\usepackage{array}
\usepackage[caption=false,font=normalsize,labelfont=sf,textfont=sf]{subfig}
\usepackage{textcomp}
\usepackage{stfloats}
\usepackage{url}
\usepackage{verbatim}
\usepackage{graphicx}
\usepackage{cite}
\usepackage{booktabs}
\graphicspath{{Figure/}}
\def\BibTeX{{\rm B\kern-.05em{\sc i\kern-.025em b}\kern-.08em
    T\kern-.1667em\lower.7ex\hbox{E}\kern-.125emX}}
\begin{document}

\title{Pose-aware 3D Beamwidth Adaptation for Mobile Extended Reality}

\author{\IEEEauthorblockN{ 
Alperen Duru\IEEEauthorrefmark{1},
Mohammad Mozaffari\IEEEauthorrefmark{2},
Mehrnaz Afshang\IEEEauthorrefmark{2},
Ticao Zhang\IEEEauthorrefmark{2},
Talha 
Khan\IEEEauthorrefmark{2},\\
Todd E. Humphreys\IEEEauthorrefmark{1},
and Jeffrey G. Andrews\IEEEauthorrefmark{1}
\vspace{0.3cm}}
\IEEEauthorblockA{\IEEEauthorrefmark{1}University of Texas at Austin, Austin, TX, USA\\ Emails: \{aduru, todd.humphreys, jandrews\}@utexas.edu}
\IEEEauthorblockA{\IEEEauthorrefmark{2}Ericsson Research, Silicon Valley, Santa Clara, CA, USA. \\ Emails: \{mohammad.mozaffari, mehrnaz.afshang, ticao.zhang, talha.khan\}@ericsson.com}
}

\maketitle

\begin{abstract}
This paper presents a sensor-aided pose-aware beamwidth adaptation design for a conceptual extended reality (XR) Head-Mounted Display (HMD) equipped with a 2D planar array. The beam is tracked and adapted on the user side by leveraging HMD orientation estimates. The beamwidth adaptation scheme is effected by selective deactivation of elements in the 2D antenna array, employing the angular estimation covariance matrix to overlap the beam with the estimation confidence interval. The proposed method utilizes the estimation correlations to adapt the beamwidth along the confidence interval of these estimates. Compared to a beamwidth adaptation without leveraging estimation correlations, the proposed method demonstrates the gain of leveraging estimation correlations by improving the coverage area for a given outage probability threshold by approximately 16\%, or equivalently increasing the power efficiency up to 18\%.
\end{abstract}


\section{Introduction}
Extended reality, which encompasses virtual, augmented, and mixed realities, demands data rates up to Gbps levels and latency as low as few tens of milliseconds. Advances towards meeting these demands in communications for XR are being pursued as a part of 5G-Advanced and are a key part of 5G-Advanced standardization. Within an XR communication framework integrated with edge computing, prolonged battery life is also a key requirement of user equipment \cite{hande2023extended}. The high data rate demands of XR require increased bandwidth, and consequently, higher carrier frequencies. Higher frequency signals face higher penetration losses and blockage \cite{hemadeh2017millimeter}, motivating the adoption of multiple antenna communication frameworks with beamforming capabilities.

Although beamforming can counteract path loss at high frequencies, it introduces a connectivity challenge for XR applications. Rapid changes in XR HMD's position and orientation complicate the precise tracking of the beamforming direction. Such inaccuracies can lead to beam misalignment, resulting in communication outages, particularly with narrow beamwidths. A narrow beamwidth makes the system susceptible to communication outages arising from headset position and orientation uncertainty, highlighting the need for a robust beam adaptation algorithm.

User-side beamwidth adaptation can mitigate beam misalignment issues stemming from high mobility and headset rotations, thereby saving energy and improving coverage. At pedestrian speeds, the primary obstacle to tracking the beamforming direction arises from inaccuracy in headset orientation estimation. The proposed user-side beamwidth adaptation scheme is based on selective deactivation of elements in the 2D antenna array to maximize overlap between the beam and the uncertainty ellipse of the desired beamforming direction.

\subsection{Related Work and Its Limitations}

Prior literature on beamwidth adaptation extensively focuses on vehicular channels where the beam is adapted at the Base Station (BS), particularly for high vehicle position uncertainty or high vehicle speed. Studies specific to user-side beamwidth adaptation, especially concerning HMDs, are scarce. This paper aims to address this gap, which is pertinent to the XR communication landscape.

Authors of \cite{va2015} explore beamwidth adaptation for high-speed train infrastructures, addressing location and velocity uncertainties. Similarly, \cite{zhu2019} and \cite{va2016} use shared user locations with BS for adapting beamwidth at the BS side. These works largely rely on 1D antenna configurations, in contrast to the 2D antenna approach adopted in this paper. A user-side beam adaptation for a 2D antenna with three degrees of freedom (3DoF) is proposed in \cite{struye2023}, where the focus is indoor environments. Authors of \cite{yang2019} consider outdoor path loss effects by adjusting the beam according to UAV's distance to BS, but the study overlooks user rotations. Many studies such as \cite{chung2021,karacora2023,peng2020,yang2015}, where \cite{chung2021} uses antenna deactivation, either address mmWave or terahertz frequency beamwidth adaptations with a focus on 1D antenna arrays or explore multi-level 3D beamforming techniques without considering sensor estimation effects on the communication outages. In \cite{tagliaferri2021}, the authors discuss the utilization of confidence intervals, with beamwidth adaptation being restricted solely to either azimuth or elevation dimensions. Authors of \cite{bell2000bayesian,vorobyov2003robust} leveraged the covariance matrices to their full extent, however, they do not address the power saving concerns of XR.

The existing body of work either gravitates towards indoor settings, permitting broader beamwidths at the cost of gain, or leans on vehicular channel assumptions, not incorporating the 3DoF capabilities or low-power requirements of XR HMDs. This underscores a need for a versatile beamwidth adaptation approach suitable for diverse environments and universally applicable to 6DoF terminals, be it a BS or User Equipment (UE).

\subsection{Contributions}
This paper offers an analysis of XR communication from the user's perspective, emphasizing path loss effects and beamwidth adaptation for 2D antenna arrays with 6DoF. The paper's primary contributions are (1) a user-side 3D beamwidth adaptation technique leveraging the 2D antenna array characteristics for outdoor scenarios and (2) analysis and simulations showing coverage distance and power efficiency improvements. 

The proposed beamwidth adaptation improves the communication coverage region by up to 16\% and the power efficiency of the receive antennas by up to 18\% compared to a beamwidth adaptation without leveraging estimation correlations. Increased power efficiency prolongs the battery life of the XR HMD, which is a key requirement for an immersive XR. This paper extends the current literature on beamwidth adaptation to the outdoor scenario for a 2D antenna array with 6DoF characteristics. The analysis considers the required array gain combined with the beam coverage area by incorporating path loss effects. The generated beamwidth adaptation algorithm is applicable to any 6DoF terminal with a 2D antenna array structure.

\section{System Model}
\begin{figure}[!t]
\centering
\includegraphics[width=0.45\textwidth]{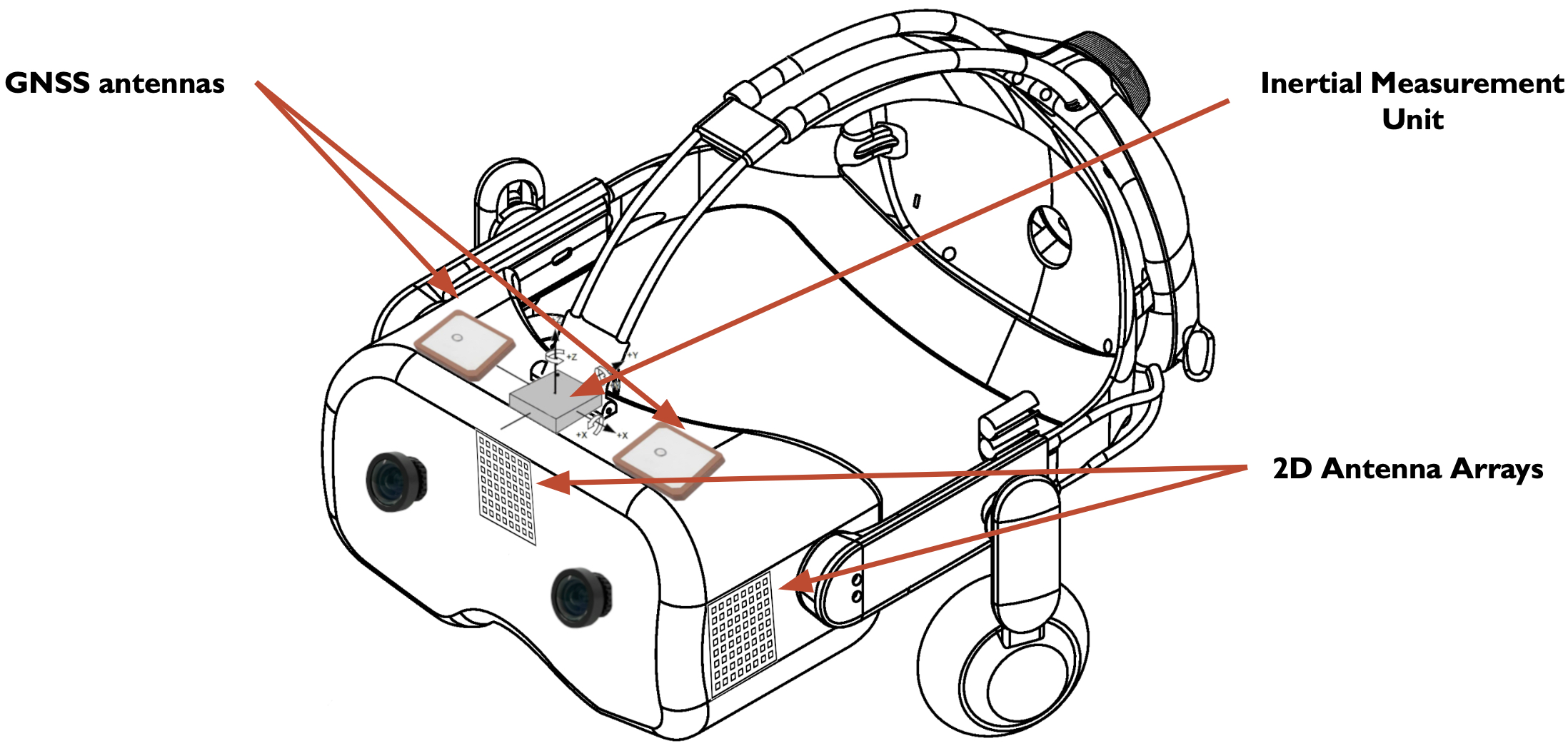}
\caption{Conceptual XR design with onboard GNSS and IMU sensors.}
\label{fig:Figure2}
\end{figure}

An XR headset communicating with the BS in downlink is considered. The headset is assumed to be equipped with dual Global Navigation Satellite Systems (GNSS) antennas and an Inertial Measurement Unit (IMU) sensor for precise position or orientation estimation. These estimates undergo filtration via an Unscented Kalman Filter (UKF), producing data on location, orientation, translational and rotational velocities, and their associated covariance matrices. The XR HMD is equipped with a 2D antenna array operating at 28 GHz, featuring constant power per antenna and the capability to deactivate individual elements, as illustrated in Fig. \ref{fig:Figure2}. Operating at pedestrian speeds of 5 km/h on a horizontal plane, the headset supports 3DoF rotation. The UE to BS distance spans 10 m to 4 km, with orientation accuracy becoming crucial for beam alignment and minimizing communication outages at greater distances.

It is assumed that the initial alignment or realignment of beamforming is established in a Line-of-Sight (LOS) link between a singular user and a BS. The XR device conveys its location to the BS via a reliable communication link different than the 28 GHz link, enabling the BS to direct its beam accurately toward the user. Given the precise shared location and limited user speed, the user's location changes only cause small angular errors for the beamforming from the BS. Thus, as a minor simplification, the beamforming from the BS is assumed to be flawless. Therefore, this paper exclusively focuses on alignment errors stemming from user orientation estimation errors.

The analysis depends on distance-dependent path loss and neglects fast and slow fading effects. Fast fading is neglected due to the assumed strong LOS link being dominant compared to low-power reflections. Shadowing is also neglected, as the angular changes due to HMD rotations are much faster than the changes due to shadowing. With these assumptions, the received signal power is modeled as

\begin{equation}
P_{\mathrm{rx}}=P_{\mathrm{tx}} G_{\mathrm{tx}} G_{\mathrm{rx}}\left(\frac{\lambda}{4 \pi d}\right)^\alpha\label{Prx},
\end{equation}
where $P_{\mathrm{rx}}$ and $P_{\mathrm{tx}}$ are receive and transmit powers, $G_{\mathrm{tx}}$ and $G_{\mathrm{rx}}$ are the beamforming gains at the transmitter and receiver, $\alpha$ is the path loss exponent, $\lambda$ is the carrier wavelength, and $d$ is the distance between transmitter and receiver. With the system assumptions, $G_{\mathrm{tx}}$ and $G_{\mathrm{rx}}(\theta, \psi)$ are defined as
\begin{equation}
    G_{\mathrm{tx}}=N_{\mathrm{tx}} \quad G_{\mathrm{rx}}(\theta, \psi)=\mathrm{AF}(\theta, \psi),
\end{equation}
where $\theta$ and $\psi$ are the elevation and azimuth angles, $\operatorname{AF}(\theta, \psi)$ is the receive beamforming array factor (AF), and $N_{\mathrm{tx}}$ is the number of transmit antennas. Due to the low mobility of the user, beamforming from the BS is assumed to be flawless within the signal reception time, and the transmit beamforming gain is thus set as the number of antennas.

The 3GPP antenna pattern is adopted in this study, where the effective area gain of a single antenna consists of vertical and horizontal, or elevation and azimuth angles, respectively. The vertical and horizontal radiation patterns are given as follows \cite{rebato2018}:
\begin{equation}
\begin{aligned}
& \mathrm{A}_{\mathrm{E}, \mathrm{V}}(\theta)=-\min \left\{12\left(\frac{\theta}{\theta_{3\mathrm{~dB}}}\right)^2, \mathrm{S L}_{\mathrm{V}}\right\} \\
& \mathrm{A}_{\mathrm{E}, \mathrm{H}}(\psi)=-\min \left\{12\left(\frac{\psi}{\psi_{3\mathrm{~dB}}}\right)^2, \mathrm{A}_\mathrm{m}\right\},
\end{aligned}
\end{equation}
where $\mathrm{SL}_{\mathrm{V}}$ is the side-lobe level limit and $\mathrm{A}_\mathrm{m}$ is the front-back ratio. The overall 3GPP antenna gain pattern for an individual antenna in terms of $\mathrm{dB}$ is given below, where $G_{\max }$ is the maximum gain in the main lobe center.
\begin{equation}
\mathrm{AP}_{3 \mathrm{GPP}}(\theta, \psi)=G_{\mathrm{max}}-\min \left\{-\mathrm{A}_{\mathrm{E}, \mathrm{V}}(\theta)-\mathrm{A}_{\mathrm{E}, \mathrm{H}}(\psi), \mathrm{A_m}\right\}
\end{equation}

The 3GPP antenna pattern is used to adjust the AF. Since the user orientation estimation errors are assumed to be the main challenge for beam alignment, the AF at the receiver should be adapted.


\section{Outage Probability}
A low outage probability is pivotal for latency, as every outage triggers a beam realignment, thereby increasing latency and reducing energy efficiency. In the XR context, where latency and energy are constrained, a beamforming algorithm resilient to estimation uncertainties is essential to preserve the user quality of experience.

An outage is declared if the received SNR falls below a certain threshold. From the array gain perspective, the SNR threshold transforms into a beamforming gain threshold $G_0$ as
\begin{equation}
    \mathrm{AF}(\theta, \psi)<G_0.
\end{equation}

The UKF models the source direction of arrival (DoA) estimation as a jointly Gaussian random variable across elevation and azimuth angles, which is characterized entirely by its covariance matrix $\Sigma_{\theta,\psi}$.
The outage probability due to misalignment can be expressed as an integral over a specific region determined by the gain threshold and AF. Let $\boldsymbol{x} = [\theta, \psi]^\mathrm{T}$, let $\boldsymbol{\mu} = [\bar{\theta}, \bar{\psi}]^\mathrm{T}$ be the mean source DoA, and define the set $\mathcal{X} = \{\boldsymbol{x} : |\mathrm{AF}(\theta, \psi)| \geq G_0 \}$. Then
\begin{align}
P_{\text{out}} =&  1 - \underset{\boldsymbol{x} \in \mathcal{X}}{\iint}\frac{1}{2\pi\sqrt{|\Sigma_{\theta, \psi}|}} e^{-\frac{1}{2} (\boldsymbol{x}-\boldsymbol{\mu})^\mathrm{T} \Sigma_{\theta, \psi}^{-1} (\boldsymbol{x}-\boldsymbol{\mu})} ~ d\boldsymbol{x} \nonumber \\
=& 1 - \underset{\boldsymbol{x} \in \mathcal{X}}{\iint} \mathcal{N}(\boldsymbol{x}; \boldsymbol{\mu}, \Sigma_{\theta, \psi}) ~d\boldsymbol{x}. \label{Pout}
\end{align}
 The integral limits, defined by $|\mathrm{AF}(\theta,\psi)|\ge G_0$, may be incorporated via an indicator function to simplify the integral in (\ref{Pout}). Since the focus of this paper is to analyze the effects of source DoA estimation inaccuracy on the outage probability, the required gain $G_0$ for a given communication distance is set to be a constant. The indicator function is defined as
\begin{equation}
 \mathrm{I}_\mathrm{AF}(\theta, \psi) = 
\left\{
    \begin{array}{lr}
      1, & \text{if } |\mathrm{AF}(\theta, \psi)| \in \mathcal{X} \\
0, & \text{if } |\mathrm{AF}(\theta, \psi)| \notin \mathcal{X} 
    \end{array}
\right..\label{Indicator}
\end{equation}
The integral effectively spans regions where the beam gain surpasses the threshold. Factors like changing distance, source DoA estimation error covariance matrix, SNR threshold, and the receive beam array factor impact the integral and, by extension, the outage probability.

\section{Beamwidth Adaptation Approach}
\subsection{General Setup and Problem Formulation}
The outage probability, once characterized by misalignment, can be managed by modifying the beam shape, which depends on the array factor and, consequently, on individual array weights. By altering these weights, the outage probability can be minimized. For a total of $N$ antennas with element weights $\boldsymbol{w}_n \in \mathbb{C}^{N}$, the array factor is
\begin{equation}
    \mathrm{AF}(\theta, \psi) = \sum_{n=0}^{N-1} \boldsymbol{w}_n e^{-j\boldsymbol{k}^\mathrm{T} \boldsymbol{d}_n},
\end{equation}
where $\boldsymbol{d}_n$ defines the element 3D location of array element n, and $\boldsymbol{k} = \frac{2 \pi}{\lambda}[\sin \theta \cos \psi, \sin \theta \sin \psi, \cos \theta]^{\mathrm{T}}$ represents the wave vector with $\lambda$ as the carrier wavelength. Traditionally, the beam direction can be controlled by adjusting the phase of each element with the element weights $\boldsymbol{w}_n$.

Optimizing the array weights maximizes the integral in the outage probability definition. The optimal array weights $\boldsymbol{W}_n$ for the optimization of (\ref{Pout}) through antenna activation becomes
\begin{align}
\boldsymbol{W}_n  = \arg\max_{\boldsymbol{w}_n}  \int_{0}^{2\pi} \int_{-\frac{\pi}{2}}^{\frac{\pi}{2}} \mathcal{N}\left(\boldsymbol{x} ; \boldsymbol{\mu}, \Sigma_{\theta, \psi}\right)
  \mathrm{I}_\mathrm{{AF}} (\theta, \psi) ~ d\boldsymbol{x}.
  \label{WnOptimization}
\end{align}

One practical approach to minimizing outage probability is activating a subset of antennas after steering the beam to the target to adapt beamwidth. Antenna deactivation is a straightforward method, particularly appealing from the user side due to its simplicity and energy-saving potential.

While other beamwidth adaptation techniques might need complex antenna structures, specific antenna couplings, and/or intricate codebooks, antenna deactivation offers a simple solution by avoiding the complexities of individually adjusting gains and phases. Given its benefits, this study adopts antenna deactivation to examine the nuances of beamwidth adaptation for XR HMDs.

\subsection{Confidence Interval Estimation}
The covariance matrix $\Sigma_{\theta, \psi}$, derived from angular errors in elevation and azimuth domains, facilitates the generation of the confidence interval. The eigendecomposition of this matrix characterizes its confidence interval orientation, width, and height, which is tied to specific probabilistic coverage requirements. Using the covariance matrix, obtained through sensor estimations and UKF filtering, the confidence interval size and orientation are characterized by its eigenvalues and corresponding eigenvectors.

The covariance matrix, acting as a linear operator on white noise, outputs colored noise, resulting in an ellipse influenced by the matrix's eigenvectors. For angular estimations, E, when projected onto a unit norm vector $\hat{a}$, the variance of this projected data, $\sigma_a^2$, is defined as $\hat{a}^\mathrm{T}\Sigma_{\theta, \psi}\hat{a}$

To pinpoint $\hat{a}$, $\sigma_a^2$ is optimized towards the direction with the greatest variance. This process is like maximizing the Rayleigh Quotient associated with the dominant eigenvector of $\Sigma$, revealing the confidence interval's orientation. The subsequent leading eigenvalues further detail the distribution's spread and orientation.

\subsection{Outage Probability Optimization}
For optimal coverage, the beam should span the entire uncertainty region which has an elliptical confidence interval due to the joint Gaussian behavior of angular estimations. By orienting the beam with the confidence interval, outage probability is optimized.

To prove this, the beam pattern is approximated for uniform gain within its 3 dB beamwidth, and zero elsewhere. This mirrors the gain threshold and path loss considerations in the outage probability's integration region. The effective array gain distributed over an area $\mathcal{S}$ can be approximated as
\begin{equation}
    |\mathrm{AF}(\theta, \psi)| \approx \frac{\pi^2}{|\mathcal{S}|}.
\end{equation}

This approximation applies to the ideal scenario, where the beam gain is equally distributed along the region of interest to maximize the angular coverage for a fixed power.

\begin{figure}[!t]
\centering
\includegraphics[width=0.45\textwidth]{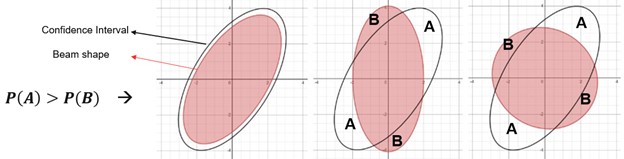}
\caption{Beam perfectly overlapping with the confidence interval (left), beam eccentricity is the same as the confidence interval, but orientation is not aligned (middle), beam eccentricity is not the same as the confidence interval (right).}
\label{fig:Figure1}
\end{figure}

Given the elliptical nature of the confidence interval and the Gaussian PDF's exponential decay, an offset in the beam orientation from the interval center leads to reduced probability coverage. Differing beam shapes, especially those not aligned with the confidence interval, yield increased outage probabilities. Then, a locally optimal beam shape aligns closely with the confidence interval, as depicted in Fig. \ref{fig:Figure1}.

Similarly, when the beam shape and thus the coverage area is different, the probability coverage will be smaller, increasing the outage probability. As a result, the optimum beam shape should be the one overlapping the confidence interval. When the beam is adapted by antenna deactivation, higher power requirement translates to more antenna activation. Optimizing coverage, therefore, involves aligning the beam with the confidence interval and adjusting the number of active antennas.

Let $\mathrm{I}_{\mathrm{AF}}\left(\phi_\mathrm{\Sigma}, N_\mathrm{A}, \theta, \psi\right)$ denote an indicator function of the beam shape aligned with the confidence interval, where $\phi_\mathrm{\Sigma}$ is the confidence interval orientation angle with respect to its azimuth estimation axis and $N_{\mathrm{A}}$ is the number of active antennas. Since optimizing the outage probability involves aligning the beam with the confidence interval and changing the number of active antennas to cover the most area possible, optimization in (\ref{WnOptimization}) can be rewritten with $\mathrm{I}_{\mathrm{AF}}\left(\phi_\mathrm{\Sigma}, N_\mathrm{A}, \theta, \psi\right)$ as

\begin{align}
    \boldsymbol{W}_n = \arg\max_{\boldsymbol{w}_n} \left[ \int_{0}^{2\pi} \int_{-\frac{\pi}{2}}^{\frac{\pi}{2}} \mathrm{I}_{\mathrm{AF}} (\phi_\mathrm{\Sigma}, N_\mathrm{A}, \theta, \psi) \, d\theta d\psi \right].
\end{align}
Once the beam is steered and aligned with the confidence interval, optimizing the outage probability becomes selecting the appropriate number of active antennas.

\subsection{Generalized Beamwidth Adaptation}
Conventionally, the beamwidth is adapted along only azimuth and elevation directions, however, utilizing the 2D antenna array structure of the XR HMD helps create a beam that aligns with the shape of the confidence interval by activating a subset of the antennas. To produce a beam matching the confidence interval's elliptical shape, the relationship between the 3 dB beamwidth and the antenna count is vital and given in (\ref{BwNumAnt}). Exploiting the covariance matrix's eigenvectors and eigenvalues facilitates the generation of the desired beam shape. The required beamwidth within the 2D azimuth-elevation domain can then be mapped to the necessary antenna count in specific directions:
\begin{align}
    b_\beta = \frac{0.886\lambda}{N d \cos \beta}.\label{BwNumAnt}
\end{align}

Here, $b_{\beta}$ is the beamwidth, $N$ is the number of antennas, $d$ is the antenna spacing, and $\beta$ is the beam pointing direction. This equation converts the confidence interval's widths defined by its covariance matrix eigenvalues $a$ and $b$ to the required antenna numbers. The confidence interval defines the required beamwidths and these beamwidths can be converted into required antenna counts $N_\mathrm{a}$ and $M_\mathrm{a}$ along their eigenvectors, with the inequalities representing antenna count boundaries:
\begin{align}
2M_\mathrm{a} &\leq \frac{0.886}{\left(  \frac{b\pi}{180} \right) \cos \beta} \\
2N_\mathrm{a} &\leq \frac{0.886}{\left(  \frac{a\pi}{180} \right) \cos \beta}.
\end{align}

The identified beamwidths across the confidence interval's dimensions construct an ellipse in the number of antennas domain. A specific antenna subset guaranteeing a minimum of 3 dB beamwidth aligned with the confidence interval is chosen. The antenna ellipse is defined as
\begin{equation}
    \frac{\left(m \cos \left(\phi_\mathrm{\Sigma}\right)+n \sin \left(\phi_\mathrm{\Sigma}\right)\right)^2}{M_\mathrm{a}^2}+\frac{\left(m \sin \left(\phi_\mathrm{\Sigma}\right)-n \cos \left(\phi_\mathrm{\Sigma}\right)\right)^2}{N_\mathrm{a}^2} \leq 1 . \label{AntennaEllipse}
\end{equation}

In (\ref{AntennaEllipse}), $m$ and $n$ denote the antenna indices and $\phi_\mathrm{\Sigma}$ signifies the confidence interval's orientation relative to the azimuth axis. The ellipse in (\ref{AntennaEllipse}) is placed at the center of the 2D antenna array and antennas with indices inside the ellipse are activated to achieve the beam. Fig. 3 depicts the transformation of a sample confidence interval into a beam aligned with that confidence interval.

\begin{figure}[!t]
\centering
\includegraphics[width=0.5\textwidth]{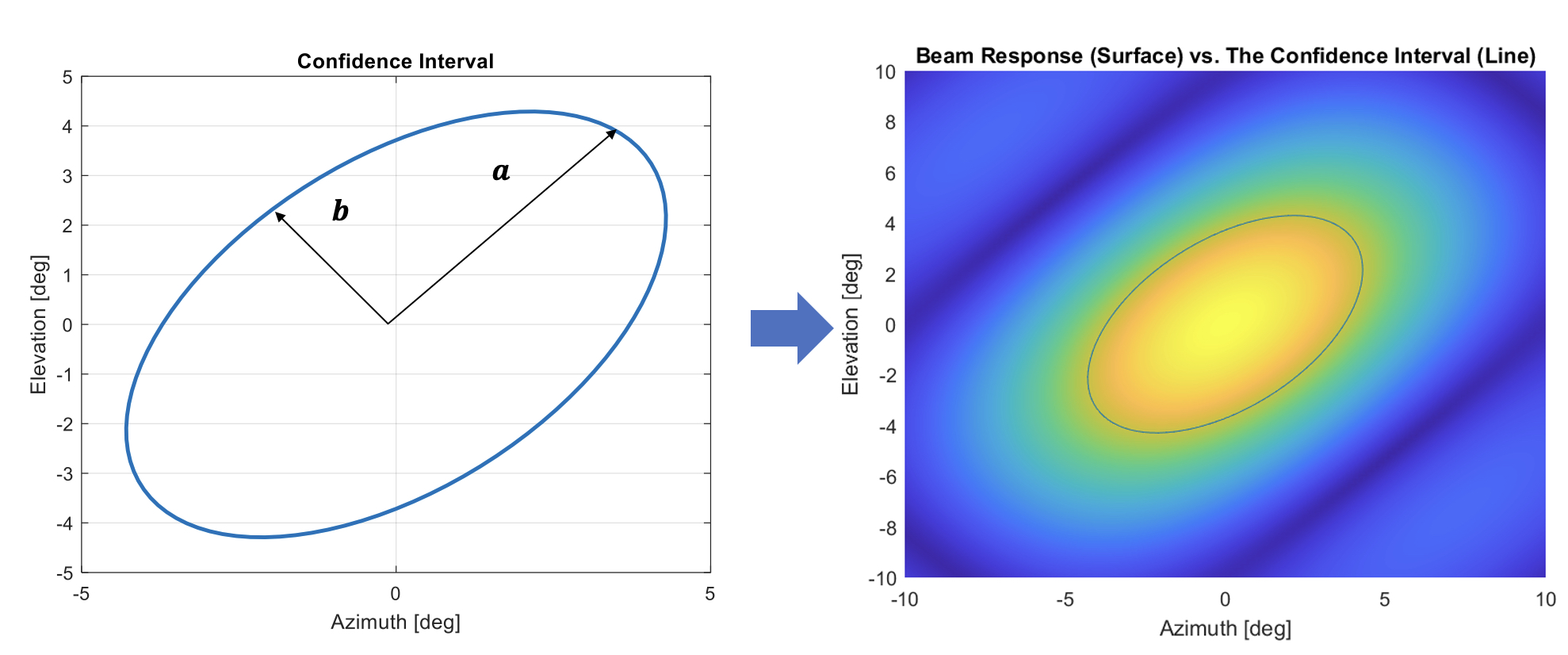}
\caption{Transformation of confidence interval (left) to the resultant beam shape (right).}
\label{fig:Figure3}
\end{figure}

\section{Results and Discussion}
The system was simulated in MATLAB for varying link distances and for the 3GPP antenna pattern with a 16 by 16 receive antenna at the XR HMD, comparing the proposed algorithm with the beamwidth adaptation with antenna deactivation without utilizing the estimation correlation. The 3GPP antenna pattern features 3 dB beamwidths of 65 degrees in both vertical and horizontal planes and a peak gain of 8 dBi at the main lobe center. Estimation correlation was assumed to study the impact of the covariance matrix on outage probability improvement. Orientation standard deviation was set at 2 degrees for both azimuth and elevation, with a correlation metric of 0.5. The SNR threshold was established at 3 dB, necessitating varied gain thresholds for different communication distances. The communication parameters are given in Table \ref{tab:parameters}.

\begin{table}[ht]
  \centering
  \caption{Communication Parameters for Simulations}
  \begin{tabular}[c]{lr}
    \toprule
    Parameter &  Value \\ \midrule
    Transmit Power (dBm/MHz) & 3 \\
    Bandwidth (MHz) & 400 \\
    Carrier Frequency (GHz) & 28 \\
    Number of BS Antennas & 256 \\
    Noise PSD (dBm/Hz) & -174 \\
    Path Loss Exponent ($\alpha$) & 2.5 \\
    SNR Threshold (dB) & 3\\ \bottomrule
  \end{tabular}
  \label{tab:parameters}  
\end{table}

\subsection{Optimal Number of Active Antennas}
The number of antennas, subject to an application-specific outage probability threshold and latency constraint, has both upper and lower bounds. The minimum number ensures the required gain, while the maximum is constrained by the inverse relationship between antenna count and beamwidth; as the number of antennas increase, beamwidth decreases, subsequently increasing the outage probability. The allowable number of antennas for a given outage probability threshold is given as
\begin{align}
    N_{\text{min}} \leq N \leq N_{\text{max}}.
\end{align}

\begin{figure}[!t]
\centering
\includegraphics[width=0.45\textwidth]{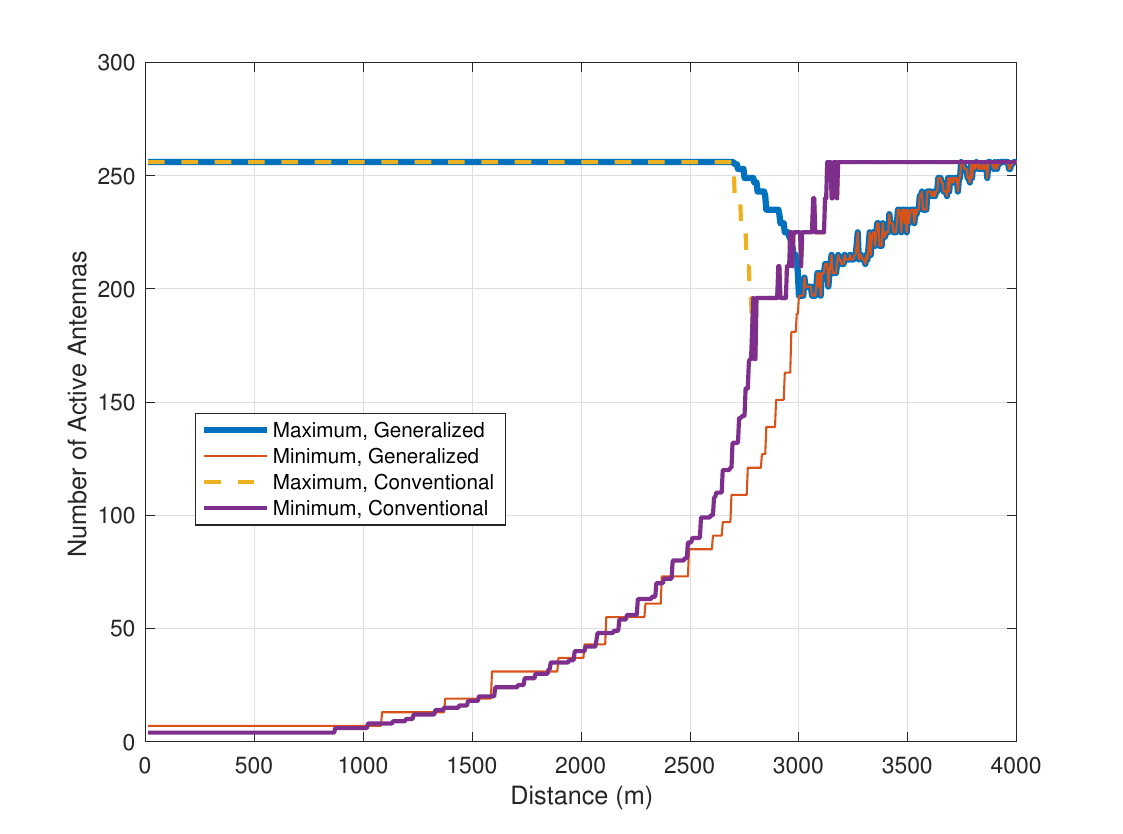}
\caption{Number of allowable antennas versus distance for the 3GPP antenna model.}
\label{fig:Figure4}
\end{figure}

Fig. \ref{fig:Figure4} depicts the range of antennas permissible for a given outage probability. Beyond a certain link distance, the maximum and minimum antenna counts align, indicating the system's effort to maintain the desired outage probability, even when it may no longer meet the outage threshold. The proposed adaptation algorithm extends coverage distance up to 8\% by utilizing the covariance matrix rather than simply adapting along azimuth and elevation. The 8\% increment in coverage distance translates into approximately 16\% coverage area increase. Variations in the correlation metric could modify this improvement, with gains approaching conventional performance in the absence of correlation.

The proposed algorithm reduces antenna usage by up to 18\% at certain distances to maintain the same outage probability, leading to equivalent power savings. Fig. 5 shows that the optimal number of antennas for minimizing outage probability lies between the maximum and minimum limits specified in Fig. 4, according to a set outage probability threshold.
\begin{figure}[!t]
\centering
\includegraphics[width=0.45\textwidth]{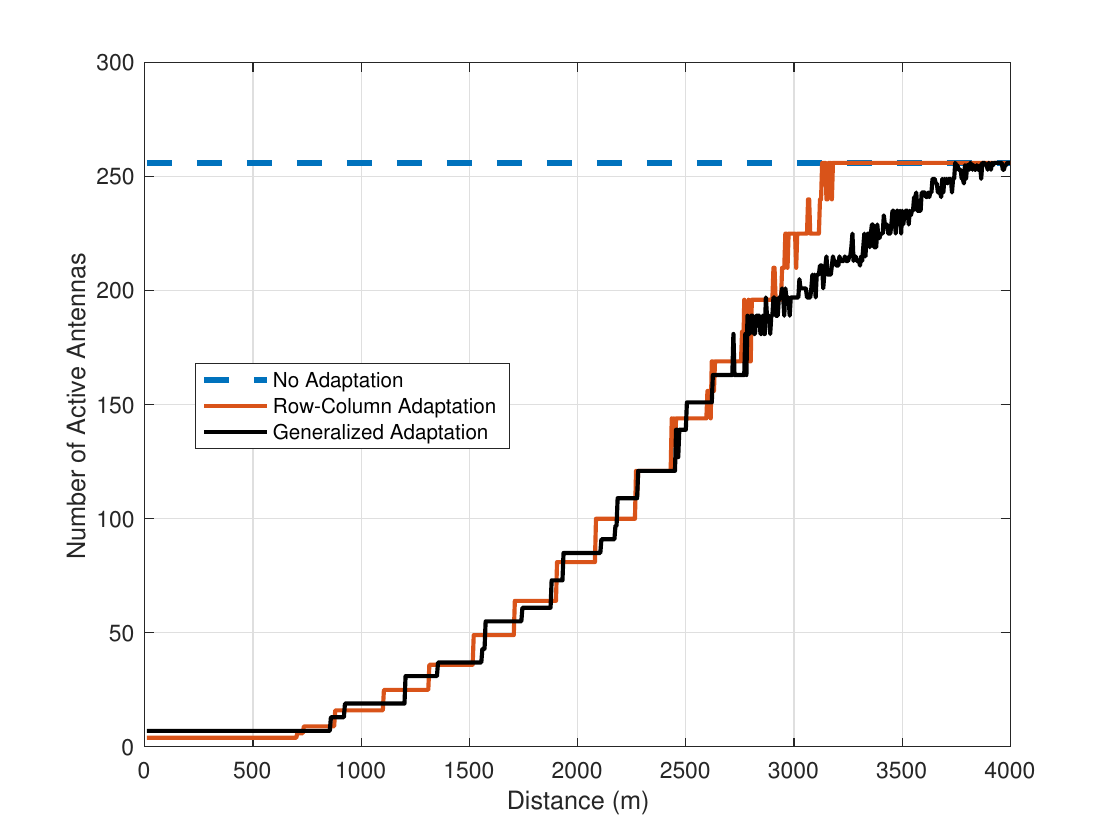}
\caption{Optimal number of antennas versus distance for the 3GPP antenna model.}
\label{fig:Figure5}
\end{figure}

\subsection{Changing Correlation’s Effect on Outage Probability}
\begin{figure}[!t]
\centering
\includegraphics[width=0.45\textwidth]{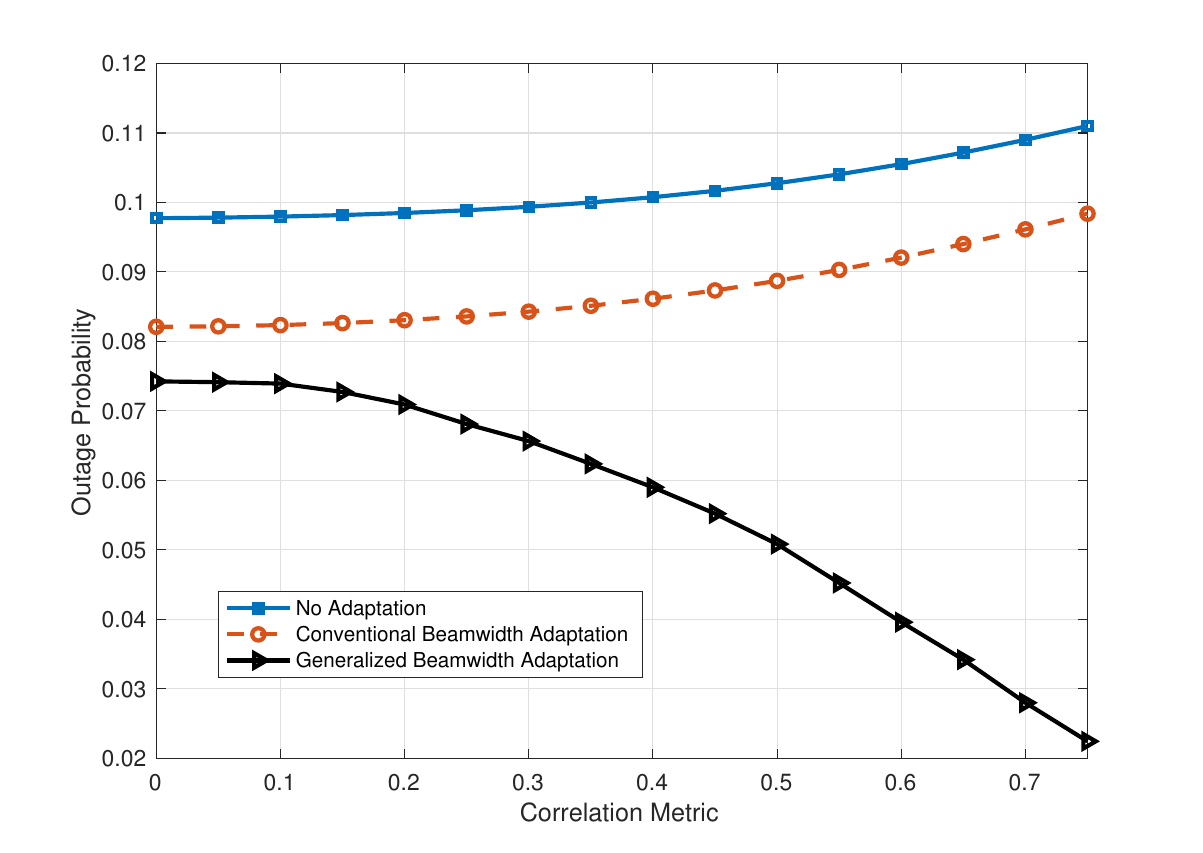}
\caption{Outage probability versus correlation metric of the covariance matrix for the 3GPP antenna model.}
\label{fig:Figure6}
\end{figure}

As the correlation metric increases, the generalized beamwidth adaptation uses the correlation information between the azimuth and elevation to its advantage. Therefore, it is expected and can be observed in Fig. \ref{fig:Figure6} that the generalized beamwidth adaptation performs better than the azimuth-elevation only beamwidth adaptation as the correlation metric increases.

On the other hand, increasing the correlation metric decreases the adapted beam's coverage over the estimation confidence interval achieved by the conventional beamwidth adaptation technique and increases the outage probability, which can be observed in Fig. \ref{fig:Figure6}.

\subsection{Confidence Interval Orientation's Effect on Outage Probability}
\begin{figure}[!t]
\centering
\includegraphics[width=0.45\textwidth]{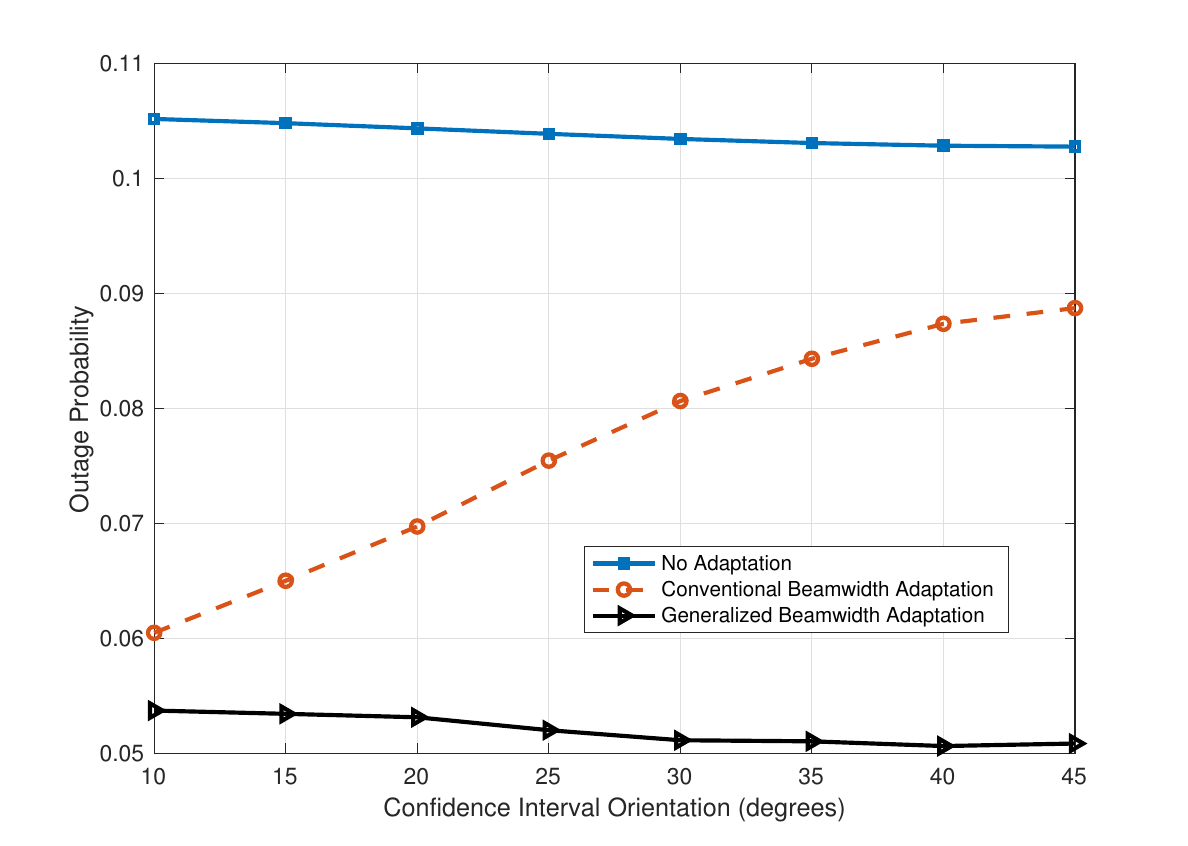}
\caption{Outage probability versus confidence interval orientation (in degrees) for the 3GPP antenna model.}
\label{fig:Figure7}
\end{figure}

When the confidence interval orientation is 0 degrees, the generalized beamwidth adaptation is essentially the same as the conventional beamwidth adaptation where only the azimuth and elevation beamwidths are adapted. As the confidence interval orientation gets closer to 45 degrees, the generalized beamwidth adaptation performs better by utilizing the covariance of the estimations. Hence, the generalized beamwidth adaptation covers the conventional beamwidth adaptation and supplies extra gain in outage probability. When the confidence interval is oriented along 45 degrees with respect to the horizontal axis azimuth, then the generalized beamwidth adaptation was observed to supply the most gain in outage probability, as depicted in Fig. \ref{fig:Figure7}.

\section{Conclusions}
In this paper, a 6DoF XR HMD communication link has been modeled to analyze the outage probability under different scenarios and link distances. Utilizing sensor covariance data enabled more effective antenna use, enhancing coverage distance by 8\% and area by over 15\%, while maintaining consistent outage probabilities and therefore, achievable latency. This increase in coverage also translates into energy efficiency, achieving comparable outage probabilities with reduced power consumption. The proposed algorithm was observed to achieve up to 18\% power efficiency improvement as a result of leveraging the DoA estimation correlations. A power savings of 18\% can substantially increase the battery life of XR HMD. Moreover, system parameter variations that affect received power levels could alter the improvement magnitude and distance at which improvements are realized.

The minimum number of allowable antennas satisfying the outage probability constraint has been observed to increase with distance to supply the required gain. On the other hand, the maximum allowable number of antennas needed to meet the outage probability threshold decreases with distance to improve the beamwidth. These two limitations meet at a critical point after which the outage probability threshold cannot be satisfied. The proposed beamwidth adaptation has pushed this critical point forward so that the coverage distance of the system has been increased. As an extension, one can define a latency metric for XR and analyze how the beamwidth adaptation can affect the upper-layer communication delays.

\bibliographystyle{IEEEtran}
\bibliography{ref.bib}

\end{document}